\documentclass[11pt,twoside,A4]{article}
\usepackage{times,fancyhdr}
\usepackage{latexsym}
\usepackage[affil-it]{authblk}
\usepackage{setspace}
\usepackage{hyperref}
\usepackage[font=small,labelfont=bf]{caption}
\usepackage[top=4cm, bottom=4cm, left=3.5cm, right=3.5cm]{geometry}
\usepackage{graphicx}
\usepackage{bm}
\usepackage{dsfont}
\usepackage{amsmath, eqnarray, amsthm, amssymb,mathrsfs}
\usepackage{xcolor}
\usepackage{soul}
\def\beq{\begin{equation}}
\def\eeq{\end{equation}}
\def\beqn{\begin{align}}
\def\eeqn{\end{align}}

\begin{document}

\title{Quantum Confusions, Cleared Up (or so I hope)}

\author{Sabine Hossenfelder}
\affil{Munich Center for Mathematical Philosophy\\ Geschwister-Scholl-Platz 1\\
D-80539 Munich, Germany}
\date{}
\maketitle

\vspace*{-1cm}

\begin{abstract}
I use an instrumental approach to investigate some commonly made claims about interpretations of quantum mechanics, especially those that pertain questions of locality. The here presented investigation builds on a recently proposed taxonomy for quantum mechanics interpretations \cite{taxonomy2023}.  
\end{abstract}

\section{Introduction}

The major problem with quantum mechanics, it seems, is that we can't agree what the problem is. Non-locality? Determinism? Realism? Incompleteness? The emergence of the classical world? All of those? None? For any possible position you could have on these points, you can find a scholar defending it. And how can we possibly solve a problem if we can't agree on what the problem is in the first place?

Underlying this disagreement about what requires our attention is a conflation of terminology, a tower of Babel that has split our people into those who speak the language of Bohmian Mechanics, and those who understand only Many Worlds, settlements of natives who grew up with a disdain for philosophy and now correspond only in maths, immigrants who are fluent in causal models but can't pronounce ``phenomenology'', experimentalists who communicate solely with grunts and grins, and nomadic iconoclasts who have invented their own language. We fight over the meaning of words, not over physics. It's what we've been doing for a century, it's not getting us anywhere, and it's about time we clean up our act.

In this paper, I want to look at some common confusions in the foundations of quantum mechanics using a purely instrumentalist approach. In particular, I want to address the questions of whether decoherence solves the measurement problem (partially, see Section \ref{sec:Dec}), whether the Collapse Postulate is necessary (it is, see Section \ref{sec:Collapse}), whether the Many Worlds Interpretation is Locally Causal (it is not, see Section \ref{sec:MWI}), and whether Bohmian Mechanics solves the measurement problem (partially, see Section \ref{sec:BM}). 

Throughout this paper, I use the convention $\hbar =1$. 

\section{Terminology}
\label{sec:Term}

Before we start, we need to define some terms that will be used later on.

\subsection{Instrumentalism}

In this paper, I will take a purely instrumentalist approach. For me, quantum mechanics is a tool that we use to make predictions for our observations, no more, and no less. 
This does not imply that we only talk about what we can observe. Certainly an instrumentalist has their eyes on explaining observations, but if non-observable entities are useful to that end, of course the instrumentalist will happily employ them.

The instrumentalist approach will allow us to study the different interpretations and representations of quantum mechanics, using the taxonomy introduced in \cite{taxonomy2023}. For this, we will need to carefully keep track of the assumptions that we make, of the input we require, and of the observations that we can predict. 

The axioms of quantum mechanics can be summarised as follows (see eg \cite{zurek2018quantum,ballentine1970statistical}):

\begin{itemize}
\item[{\bf A1:}]\label{A1}
    The state of a system is described by a vector of norm 1 in a Hilbert space $\mathscr{H}$. We will denote it with  $|\Psi\rangle$.

\item[{\bf A2:}]\label{A2}
    Observables are described by Hermitian operators $\hat O$ and possible measurement outcomes are these operators' eigenvalues, $O_I$. We will denote the eigenvector of the eigenvalue $O_I$ with $|O_I\rangle$.

\item[{\bf A3:}]\label{A3}
    In the absence of a measurement, the time-evolution of the state is determined by the Schr\"odinger equation $i \partial_t |\Psi \rangle = \hat H |\Psi \rangle$.

\item[{\bf A4:}]\label{A4Collapse}
   After a measurement, the system is in one of the mutually orthogonal eigenvectors of the measurement observable $|O_I\rangle$. This axiom is often called the ``Collapse Postulate''.
\item[{\bf A5:}]\label{A5Born} 
    The probability of obtaining outcome $O_I$ is given by $|\langle \Psi | O_I \rangle|^2$. This axiom is known as ``Born's Rule''.\footnote{Note that the normalization of the state is necessary for this probability distribution to be well-defined.}

\item[{\bf A6:}]\label{A6}
    The state of a composite system is described by a vector $|\Psi\rangle$ in the tensor product of the Hilbert-spaces of the individual systems.
\end{itemize}

This list of axioms is not sufficient to make predictions. To do that, one also needs to at least define the Hamiltonian\footnote{Or suitable approximations. For example, actions of logical gates or optical components (such as beam splitters or non-linear crystals etc) are usually described by applying unitary transformations that fulfil the same function as the Hamiltonian operator, namely, to encode the time-evolution.} and the operator that corresponds to the variable that one wants to measure. However, we will in the following not need to talk about these, so we leave them aside here and just imagine that the system is sufficiently specified.

Once one has all those axioms in place, one inputs an initial state into the mathematical machinery of quantum mechanics. The output is then a final state, and from that---using Born's Rule---a prediction for the probability of a particular measurement outcome. In the terms of \cite{taxonomy2023}, our list of axioms is a calculational model. For the rest of this paper, I will refer to this particular calculational model as the Copenhagen Model. 

Following the terminology of \cite{taxonomy2023}, the Copenhagen Interpretation is then the set of all axiomatic formulations that are mathematically equivalent to the Copenhagen Model, and Standard Quantum Mechanics is further the set of all physically equivalent sets, that is, those who result in the same predictions for observables.

The Copenhagen Model is neither the best nor only approach to quantum mechanics. To begin with, its axioms could be rewritten into many other, mathematically equivalent, sets. A simple example might be that in the Heisenberg picture the state has no time evolution, but instead the operators do. 

It is also quite possibly the case that there are axiomatic approaches to quantum mechanics which simplify the above axioms. This is because it has been pointed out previously (see eg \cite{hardy2001quantum} and \cite{aaronson2004quantum}) that the probabilistic calculus we use in quantum mechanics is in some sense the simplest such calculus we could put onto a vector space. 
This is an interesting thought, but for our purposes it is not all that important. I here use the formulation with axioms {\bf A1} -- {\bf A6} just because I believe it is one most readers will be familiar with, and it will be sufficient to make my points. 

There are a few approaches to derive Born's Rule from other assumptions, some more, some less convincing \cite{gleason1975measures,deutsch1999quantum,hardy2001quantum,aaronson2004quantum,dakic2009quantum, vaidman2011probability,sebens2018self,hossenfelder2021derivation}. While these are certainly worthwhile, we won't need them here either. For our purposes we just need to note that there is no known derivation of Born's Rule from the other 5 axioms on our list. To date, all known derivations require other assumptions in addition.  

\subsection{Local Causality}

To make statements about the properties of a quantum mechanical interpretation, one needs more than just the above listed axioms. Most importantly, to talk about notions of locality and causality, one needs notions of space and time to begin with. It is, however, arguably the case that we describe many of our observations in terms of location and duration. For this reason, all interpretations of quantum mechanics that aspire to describe the totality of our observations must somehow account for space and time. 

We will therefore for the remainder of this paper assume that we are dealing with an interpretation that has such a notion of space and time. However, as pointed out in \cite{taxonomy2023}, there are many models for quantum mechanical processes that deal only with e.g. spin variables and cannot be said to be local in any meaningful way because they have no notion of space or anything resembling space. 

\begin{figure}
    \centering
\includegraphics[width=0.6\linewidth]{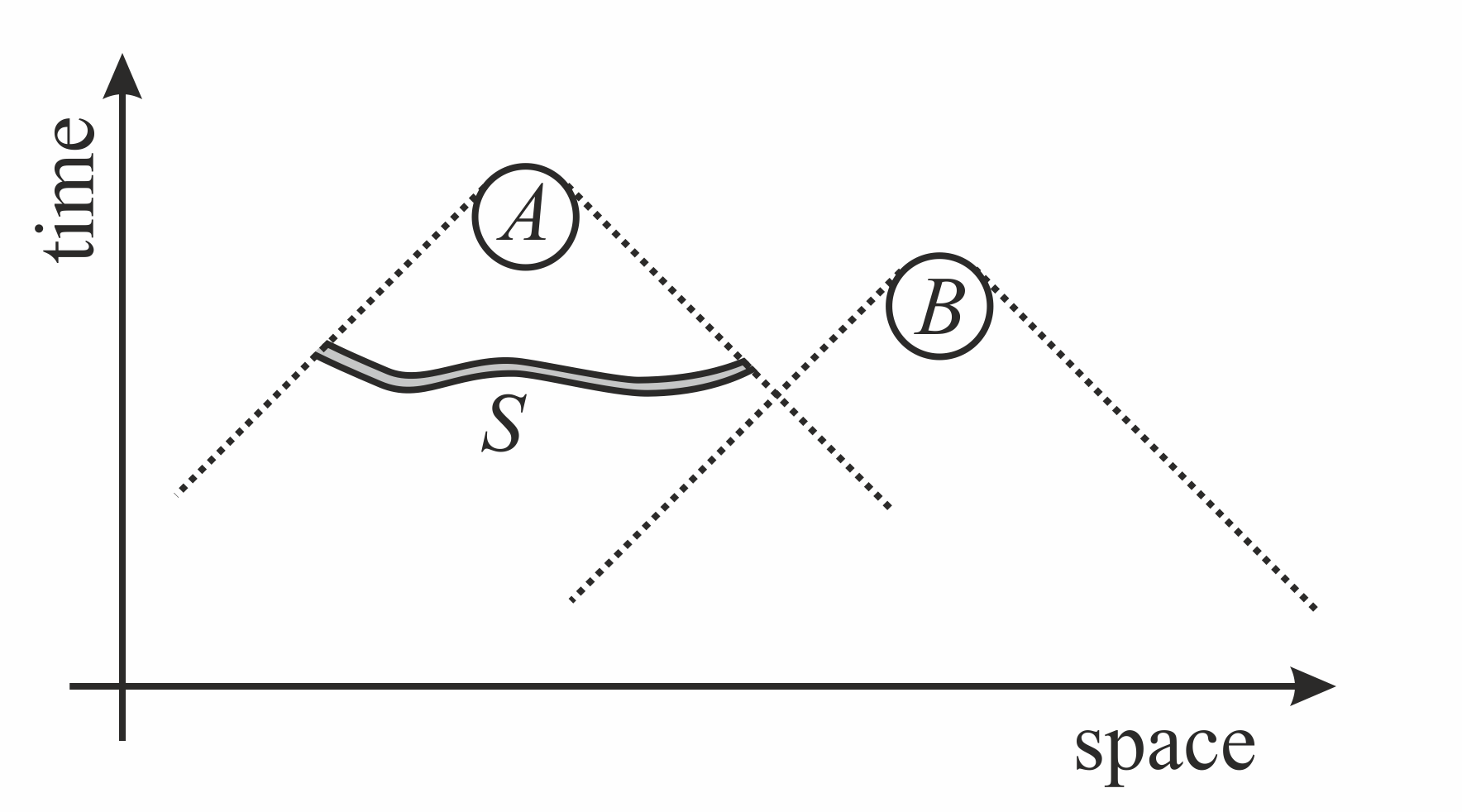} 
    \caption{Local Causality means that given complete information on $S$, what happens at $B$ tells us nothing new about what happens at $A$. The Copenhagen Interpretation does not have this property.}
    \label{fig:subtlety}
\end{figure}

The most widely accepted notion of locality in quantum foundations is Bell's Local Causality \cite{bell1975theory} (see also \cite{taxonomy2023} for a definition). The Copenhagen Model---and because of that, the Copenhagen Interpretation---is not locally causal in Bell's sense. This is because making a measurement in one location---let's call it $B$---can reveal information about what happens at another, space-like separated, region $A$, and that information was not contained on all hypersurfaces ($S$) in the backward lightcone of $A$ where they did not overlap with that of $B$. In mathematical terms, Local Causality requires $P(A | S, B) = P(A | S)~~ \forall~~ S$, where $P$ denotes probability (see Figure \ref{fig:subtlety}, after \cite{bell1975theory}). 

The reason for this failure of Local Causality is that the wave-function does not contain sufficient information to calculate the measurement outcome at $A$, combined with the fact that the outcome at $B$ tells us something about the outcome at $A$. For an excellent summary of what Local Causality does and does not mean, see \cite{norsen2009local}.

The key reason quantum mechanics is non-local is therefore that the wave-function does not contain the entire information to predict the measurement outcome at $A$. The need to introduce the Collapse Postulate is a {\emph{consequence}} of this: Once one has made a measurement at $B$, the information about the outcome must be included into the description of the system, otherwise the information is incomplete and the model is unable to correctly predict conditional probabilities. 

Local Causality, defined in this way, is a property of a (mathematical) model, it is not a priori also a property of an (assumed to exist) underlying reality, and it is indeed perfectly suited for instrumentalists. Bell himself expressed this sentiment in \cite{Bell1964OnEPR}:

\begin{quote}{\emph{``I would insist here on the distinction between analyzing various physical theories, on the one hand, and philosophising about the unique real world on the other\dots I insist that [Bell's theorem] is primarily an analysis of certain kinds of physical theory.''}}
\end{quote}

Understanding that non-locality is a property of a model, we see that the non-locality of the Copenhagen Interpretation can be remedied by positing the existence of unknown information---usually called the ``hidden variables''---that are added to $S$ (see Figure \ref{fig:subtlety}), propagate locally to $A$, and thereby make the information from the space-like separated measurement $B$ redundant for calculating the outcome of a measurement at $A$. This, in a nutshell, was the {\sc EPR} argument \cite{Einstein1935EPR}, according to which the requirement of locality implies that quantum mechanics is incomplete. 

We now know from Bell's theorem \cite{Bell1964OnEPR} that any such model with hidden variables which returns us to Local Causality must violate Measurement Independence, but that's another story which has been told elsewhere \cite{Hossenfelder2020Rethinking,Hossenfelder2020Perplexed,hance2022ensemble}.

\subsection{Bell's Theorem}

Since Local Causality is a property of a model, observing violations of Bell's inequality cannot tell us anything about nature itself, but merely about the properties of the model that we use to describe nature. As worthwhile as Bell's theorem is, its implications are often overstated. For example in \cite{tumulka2021bohmian} we can read:
\begin{quote}
    {\it ``Bell's theorem actually shows that the observed probabilities in
certain experiments (Einstein-Podolsky-Rosen-Bell experiments) are incompatible with
locality, so our world must be non-local, and every theory in agreement with experiment
must be non-local.''}
\end{quote}
The correct statement would have been: ``Bell's theorem shows that the observed probabilities in certain experiments are incompatible with the assumptions of Bell's theorem. This means that any model which fulfils Local Causality and is in agreement with observation must violate Measurement Independence.''

I want to stress that this is not a matter of interpretation, it is merely a question of what can be concluded from observations and the mathematics that (correctly) describes these observations. A theorem about a certain class of models---those which fulfil the assumptions of the theorem, that is, Local Causality and Measurement Independence---can only tell us something about those models.

\section{Does Decoherence Solve the Measurement Problem?}
\label{sec:Dec}

Let us start with the question of whether decoherence solves the measurement problem because that will illustrate why the instrumentalist approach is helpful.

The measurement problem has many aspects \cite{maudlin1995three,leggett2005measprob,hance2022does}. We will for illustration here focus on two of them. First, there's the question of what a measurement is (beyond the "we know it when we see it" take) and second, whether we can get rid of the non-locality of the measurement process.

Decoherence is often described as the result of interactions with an environment that lead to a loss of quantum properties, mathematically captured by the transition from a pure to a mixed state. For example, according to Zurek ``Environment can destroy coherence between the
states of a quantum system. This is {\emph{decoherence}}.'' \cite{Zurek2003} (emph original).

However, interactions with the environment---as all interactions in quantum mechanics---are unitary transformations and as such cannot in and of themselves transform pure to mixed states. This is why, in decoherence approaches one follows up on the increase of entanglement with the operation of mathematically ``tracing out'' (basically: summing over) that part of the system which was designated as the environment. Tracing out is usually not interpreted as a physical process but as reflecting the experimentalist's lack of knowledge.

Since we take an instrumentalist approach, however, we do not care about interpretations. We simply note that if we want to use traces to calculate probabilities, then we need to add this procedure to the axioms. And then we will need further assumptions about just what to trace out. Indeed, the notion of entanglement itself makes no sense unless we divide up the system into at least two partitions. That decoherence does not resolve this aspect of the measurement problem has been discussed previously in more detail in \cite{Schlosshauer2004Decoherence,Kastner2014Einselection}.

To put this problem in somewhat simpler terms, suppose that I give you a wave-function that I tell you contains a prepared state and all the particles of the detector and their environment, but I don't tell you which part is which. To answer the question of what a measurement is, you would have to know whether this combined system is going to perform a measurement and if so, of which variable. Saying ``decoherence'' does not suffice to answer this question. 

There have been many attempts to try and find a definition that would break up a generic state into the prepared state (that one wants to measure), the environment, and the pointer states of the detector. If one knew how to do that, then one would know what to trace out, and decoherence could be used to identify the measurement variable and the circumstances under which to apply the Collapse Postulate. I would say that as to date none of these approaches have been entirely convincing, and somewhat vaguely conclude that while decoherence has the potential to solve this aspect of the measurement problem, it hasn't yet been satisfactorily done.

Decoherence also certainly does not return us to Local Causality. Even in a decoherence-based approach there remains the undeniable fact that making a measurement in one place can instantaneously tell us something about the probability of getting a measurement outcome elsewhere. It doesn't matter whether one expresses this increase of knowledge through the Collapse Postulate or through updating the meaning of entries in the density matrix to conditional probabilities.

It is also clear that decoherence cannot result in a local collapse process: If the interactions that create entanglement are to be local, then they can start at the earliest when the prepared state reaches the detector. At that point, however, it is simply too late for the wave-function to locally go into an eigenstate.  (One can imagine cases where decoherence starts earlier because particles do not propagate through vacuum, but this generically does not work.) 

This should not be surprising because we already saw earlier that the reason why the Copenhagen Model violates Local Causality is not the Collapse Postulate per se. It is rather the lack of information in the wave-function together with the fact that an observation in one location can tell us something about the measurement outcome in another location. Describing this observation is what requires some non-local update, regardless of how one wants to express it mathematically. The only thing we can do is to prevent this non-locality from being a fundamental property of reality (rather than just a property of a model which describes knowledge) by introducing hidden variables which can locally transport the information.

This is not to belittle the achievements of the decoherence program. Studying decoherence has been extremely useful in experimental settings, e.g. to understand the effects of the environment \cite{schlosshauer2019quantum}.

Some physicists have tried to ``solve'' the measurement problem by simply redefining the Collapse Postulate to mean that a measurement results in a mixed density operator. A prominent example of this confusion can be found in \cite{ArkaniHamed}. Needless to say, taking something else and naming it ``Collapse Postulate'' does not solve any problems. 

\section{Is the Collapse Postulate Necessary?}
\label{sec:Collapse}

It is sometimes questioned whether the Collapse Postulate is even necessary (e.g. \cite{zurek2018quantum}). Especially defenders of the Many World's Interpretation and its variants seem to think that the Collapse Postulate can be discarded and the other axioms are sufficient. 

Before we get into this, I want to briefly elaborate on a subtlety of the axiomatic formulation. If one formulates Born's rule by saying that $|\langle \Psi | O_I \rangle|^2 $ is the probability of finding the system in eigenstate $|O_I\rangle$ when one measures it, then one no longer needs the Collapse Postulate. This is because this formulation implicitly already says that when we measure the system, then is in an eigenstate of the measurement variable, because that's what we are calculating the probability of. This was done e.g. in \cite{zurek2018quantum}, but not in \cite{ballentine1970statistical}. I am mentioning this just to point out that sometimes Born's Rule and the Collapse Postulate are conflated with each other.

That said, it is correct of course that we can remove the Collapse Postulate from the axioms. But we then need to add some other instructions for what to do in case of a measurement. We cannot simply do nothing, for imagine I send a particle through a beam splitter and make a measurement on one path. I do not find the particle on this path, and make a second measurement on the same path. The conditional probability for not finding the particle in the second measurement, given that it wasn't there in the first measurement, should be zero. This requirement is commonly stated as saying that immediate repetition of a measurement yields the same outcome (see e.g. \cite{zurek2018quantum}).

In the Copenhagen Interpretation this conditioning on the known result is taken into account by updating the state---i.e. by using the Collapse Postulate---to reflect the increase of knowledge gained from the previous measurement. One doesn't need to do it this way, but one has to do something. For consider that we, as an instrumentalist want to know how to write a computer code to give us correct predictions. If we just apply Born's rule repeatedly without updating the state, then it will give us wrong probabilities for measurement outcomes. So we either have to update the state or our interpretation of what the state describes (or something else).

I believe the reason this fact sometimes becomes muddied up is that the Collapse Postulate describes exactly how one would {\emph{expect}} conditional probabilities in a statistical ensemble to behave, so it seems obviously true and almost unnecessary to state. However, the common axiomatic formulation of quantum mechanics is not about statistical ensembles and therefore does not contain this information even implicitly. Including it explicitly is the idea behind the statistical interpretation \cite{ballentine1970statistical}.

\section{Is the Many Worlds Interpretation Local?}
\label{sec:MWI}

Understanding the necessity of the Collapse Postulate (or equivalent) is also useful to putting the Many Worlds Interpretation into perspective, which we will turn to now. 

\subsection{Many Worlds Recap}

There are many Many Worlds Interpretations \cite{everett1957relative,dewitt1970quantum,zeh1970interpretation,deutsch1985quantum,hartle1991spacetime,  lockwood1996many,rovelli1996relational,saunders2008branching,carroll2014many,brassard2019parallel,raymond2021local,list2023many}, but they have one thing in common: The idea that it is possible to remove the Collapse Postulate from the axioms of quantum mechanics. In the Many Worlds approach one posits that all possible outcomes of an experiment happen, each in its own universe, it's just that we only ever experience one possible outcome. The process in which the other possible outcomes become inaccessible to us is referred to as the ``branching'' or ``splitting'' of worlds.

As we just saw, one cannot just discard the Collapse Postulate. For this reason, if one does calculations in any Many Worlds approach, one must effectively replace the Collapse Postulate by other assumptions, notably about what happens to observers and/or measurement devices when worlds branch. Suitably done, the outcome for the measurement probabilities is then the same as in the Copenhagen Model.

Unfortunately, the new assumptions which must be added to make the Many Worlds Interpretation work are often not explicitly stated but implicitly appear in elaborations on what an observer is. To make a long story short, the relevant property of an observer in a Many Worlds Interpretation is that they can still can only see one outcome of an experiment. This property cannot be derived from axioms {\bf A1} -- {\bf A5}. If it was only for those assumptions then any object (including observers) would deterministically split into superpositions with any experiment for which the forward evolution of the initial state is not an eigenstate of the measurement operator. The Many Worlds Interpretation therefore must sneak back in the Collapse Postulate somehow. Usually this is done by implicitly assuming that an observer is not something that exists in multiple branches at the same time. 

A typical example of how this happens can be found in \cite{carroll2014many} where we can read:
\begin{quote} {\emph{``Because these environment states are and
will be orthogonal [...] each
component evolving in accordance with the Schr\"odinger equation as if the other were not present.
They can thus be treated as separate worlds (or collections of worlds) since they are effectively
causally isolated and within each of them there are versions of Alice having clear and determinate
experiences.''}}
\end{quote}
In this paragraph, the authors of the paper use decoherence (that, as usual, draws on a pre-defined environment) to motivate a statement about mathematically ill-defined terms like ``Alice'' and her ``experience''. It remains unexplained why the time-evolution of an object once called Alice is later no longer just Alice but multiple versions of something also called Alice. (According to the authors, {\emph{``This is a
tricky metaphysical question upon which we will not speculate.''}}) 
The rest of the paper is then dedicated to developing a rule to update probabilities in the Many Worlds Interpretation, so that they agree with the Collapse Postulate of the Copenhagen Model. Well done.

I would like to make clear that I have no issue with the way that Carroll et al interpret what is effectively the Collapse Postulate. My point is merely that it does not mathematically follow from the Schr\"odinger equation and the other 4 axioms, excluding the Collapse Postulate. If the Collapse Postulate would mathematically follow from the other 5 axioms of the Copenhagen Model, then it would do so regardless of how one interprets the mathematics. It does not. Hence, if we want a mathematical procedure with the same result as the Collapse Postulate, that will require new assumptions. If one does not make those extra assumptions, one has a theory that does not make correct predictions. 

Once one equips the Many Worlds Interpretation with suitable assumptions to reproduce the Collapse Postulate, it makes the same predictions as the Copenhagen Model. Because of the extra assumptions about observers, it is however not mathematically equivalent. In the terms of \cite{taxonomy2023}, therefore, it is not a representation of the Copenhagen Model, but a physically equivalent interpretation, and can be understood as standard quantum mechanics.

The Many Worlds Interpretation is temporally deterministic, again using the terms of \cite{taxonomy2023}. This means that the final state can be predicted with certainty from the initial state. Determinism is reestablished by positing that all possible outcomes of an experiment happen, we just only see one, whose outcome we can then not predict. The Many Worlds Interpretation is therefore not predictive, again in the terms of \cite{taxonomy2023}. We note that this has nothing to do with quantum theory in particular---any non-deterministic theory could be converted to a deterministic yet non-predictive theory in this way, it is just that the need usually does not arise. 

\subsection{Locality in the Many Worlds Interpretation}

It follows from the above that the Many Worlds Interpretation violates Local Causality in exactly the same way as the Copenhagen Model. This can be seen as follows. 

Remember that the reason the Copenhagen Model violates Local Causality is that a measurement in one location can reveal information about a measurement in another, space-like separated region, and that this information could not be obtained from the wave-function.

Well, the only information we have in the Many Worlds Interpretation is also that in the wave-function, and for what our observations are concerned, it gives the same predictions as the Copenhagen Model. Hence, it also violates Local Causality. You can believe in as many other universes as you wish, making a measurement on one end of the wave-function will de facto reveal something about the outcome on the other end. If you, for example, know that a particle was measured in one place, you know it wasn't measured in another place. This, combined with the impossibility of predicting the measurement outcome from the wave-function alone, makes the Copenhagen Model non-local, and the Many Worlds Interpretation changes nothing about that.

I believe the reason that Many World supporters are confused about this point is the following. If you think that the origin of violations of Local Causality in the Copenhagen Model is the Collapse Postulate, and you discard of the Collapse Postulate, then what's left should be local. 

However, the non-locality of the Copenhagen Model does not come from the Collapse Postulate per se, it comes from the lack of information about measurement outcomes in the wave-function, combined with the fact that making a measurement in one place reveals something about another place. The Collapse Postulate is just how this increase in information is taken into account in the Copenhagen Model. The Many Worlds Interpretation doesn't change anything about the origin of the violation of Local Causality, hence it violates Local Causality the same way. It does not help that, as pointed out in \cite{waegell2020reformulating,drezet2023elementary}, the word ``locality'' in the context of Many Worlds Interpretations has been used with different meanings.

Neither does it help to add more complicated stories about how the branching occurs. The local branching idea (of which, again, there are several variants \cite{wallace2012emergent,sebens2018self,mcqueen2019defence}) has it that the splitting of the worlds obeys the speed-of-light limit. For this reason, a measurement in $B$ in Figure \ref{fig:subtlety} does not split the world in $A$ instantaneously. It is only when the two come into causal contact and observers from both places can compare their measurements that the split must have been completed. 

But for us instrumentalist this story is irrelevant. We take the measurement results from $A$ and $B$ and apply our definition of Local Causality. It is violated, and that's that. Whether you think there were multiple different universes with different outcomes at $A$ before we compared both records is irrelevant for the fact that the records which we do have in our universe show these correlations. And since there is not enough information in the wave-function to predict $A$ without the information from $B$, Local Causality is violated. Stories about local branchings do not matter. (Unless the branching contains additional information about the measurements that will be made, in which case Local Causality can be restored by violating Measurement Independence instead.)\footnote{One could try to interpret the local branching idea as an attempt to explain why a lack of Local Causality is not something to worry about, but that is not the question we are addressing here and it is also not an argument I have encountered from Many Worlds adherents.}

One way to proceed is to just argue that Local Causality is not a relevant or even meaningful criterion to require of a theory with many worlds. This is the avenue pursued in ideas like the Parallel Lives Interpretation \cite{brassard2019parallel,raymond2021local}. This is a very valid approach to pursue because indeed giving up Local Causality is one of the logical options. It still leaves one with the question of how to reconcile the measurements that we observe with the locality requirements in Einstein's General Relativity, but then some might not find this as important as I do.

If one, however, makes the mistake of thinking that the Many Worlds Interpretation is locally causal, then one has another problem, which is Bell's theorem. According to Bell's theorem, a theory can only be locally causal if it introduces hidden variables which violate Measurement Independence. The Many Worlds Interpretation does not do that, so what gives? 

Hence emerged the idea of an additional assumption in Bell's theorem which is that measurements have only one definite outcome. It is sometimes called the ``one world'' \cite{list2023many} assumption,``single world'' assumption \cite{frauchiger2018quantum}\footnote{The term was removed from the paper in the published version but can still be found in the first arXiv version}, the ``definite outcome'' assumption \cite{waegell2020reformulating}, or the ``absoluteness of observed events''\cite{schmid2023review}. This peculiar idea has it that the Many Worlds Interpretation evades the conclusion of Bell's theorem by allowing more than one measurement outcome, each in a separate universe. 
 
But all that throwing out the assumption of a definite outcome does is render Bell's theorem useless. Bell's theorem is a statement about correlations that we observe. It is a fact, not merely a mathematical assumption, that we observe only one outcome of an experiment. If one removes this assumption, then one is left with a theorem about something that does not describe our observations.

My experience with Many Worlds enthusiasts is that they follow a motte-and-bailey approach. For those not familiar with this logical fallacy, it consists of first building up an easy to defend argument (the motte) and then using mutual agreement on that easy argument to claim victory on a different argument (the bailey). It's an example of what is also known as the bait-and-switch tactic. 

The argument of Many Worlds defenders goes like this. First, they will claim that discarding of the Collapse Postulate creates a model which is axiomatically simpler than the Copenhagen Model. Dropping this postulate moreover removes the non-local element of the dynamical law. And this is both correct, but fails to mention that removing the Collapse Postulate leaves one with a model that may be simple and local, but generically gives wrong predictions.

Then happens the switch, in which the Many Worlds fan needs to add new assumptions to make the same predictions as the Copenhagen Model does. It is highly questionable that the total of their new assumptions is any simpler than the Collapse Postulate, but this is not an argument I want to make here\footnote{It would require us to first define computational complexity and this is somewhat off-topic.}. However, any set of assumptions that leads to the same observational outcome as the Collapse Postulate without also equipping the model with more information about the measurement outcome than the wave-function has, will inevitably also violate Local Causality.

The trick of Many Worlds defenders is now to first get you to agree to take their bait argument---a model which may be simple and local but does not make correct predictions---and then claim you agreed to their elaborations on different versions of Alice. 

All of this is not to say that there is something wrong with the Many Worlds Interpretation (in the bailey-version equipped with sufficient assumptions to produce correct predictions). As an interpretation of quantum mechanics it is not necessarily any worse than the Copenhagen Model. Though personally I think using the Collapse Postulate is simpler than bothering with splitting worlds and such. I would recommend to anyone who thinks that the Many Worlds Interpretation is simpler than using the Collapse Postulate that they read this paper \cite{carroll2014many}, track all assumptions, and decide for themselves.

It is indeed possible to formulate some variants of Many Worlds models that are local, but these violate Measurement Independence. This should not be surprising because this follows from Bell's theorem. For a discussion of examples, see \cite{waegell2020reformulating}. Note that in the example for a local version of Many Worlds in the last section of \cite{waegell2020reformulating}, the branching depends on the measurement settings, and hence violates Measurement Independence---as one expects. 

\section{Does Bohmian Mechanics Solve the Measurement Problem?}
\label{sec:BM}

\subsection{Bohmian Mechanics Recap}

Bohmian Mechanics \cite{Bohm1952Bohm1,Bohm1952Bohm2} uses the Schr\"odinger equation ({\bf A3}) for the wave-function $\psi$ in configuration space
\begin{eqnarray}
{\rm i} \frac{\partial \psi (\boldsymbol{q},t)}{\partial t} = - \sum_{k=1}^N \frac{1}{2 m_k} \Delta_k \psi (\boldsymbol{q}, t) + V(\boldsymbol{q},t)\psi(\boldsymbol{q},t)~,
\end{eqnarray}
where ${\boldsymbol{q}}=(q_1,q_2\dots q_N)$ are spatial coordinates, $V$ is a potential, and $m_k$ are the particle masses.
So far, this looks like ordinary quantum mechanics. But then one posits that the state of a system is described by $N$ particles whose actual position is described by $\boldsymbol{Q}=\{Q_1(t), Q_2(t) \dots Q_N(t)\}$. The particles are assumed to have velocities given by
\begin{eqnarray}
m_k \frac{dQ_i}{dt} = {\rm Im} \frac{\psi* \nabla_k \psi}{\psi^* \psi} (\boldsymbol{Q}(t),t) ~. \label{eq:Qeq}
\end{eqnarray}

In Bohmian Mechanics, one reformulates Born's Rule ({\bf A5})  into the ``quantum equilibrium hypothesis'', according to which the probability distribution of the particles described by $\boldsymbol{Q}$ is of the form $|\psi(\boldsymbol{q},t)|^2$. One can then show that if this equilibrium hypothesis is fulfilled at one time, it will be fulfilled at all times. One hence only needs it as an assumption about the initial state. One further assumes that the state of the system is really one particular value of $\boldsymbol{Q}$, though one does not know which. $\boldsymbol{Q}$ serves the role of the hidden variables.

The picture that Bohmian Mechanics offers is that the wave-function of the Copenhagen Model serves as a guiding field for point-like particles. The indeterminism of quantum mechanics arises from our lack of knowledge about the exact initial state of the Bohmian particle (or particles, if there are several). 

Bohmian Mechanics then does not need the Collapse Postulate. This is because introducing the new variables $Q_i$ and positing that the state is only in one of the possible configurations, means that a measurement merely reveals which state the system was really in. The pilot wave itself is not updated upon measurement. This gives rise to what has been dubbed ``empty waves'' \cite{hardy1992existence,deotto1998bohmian,lewis2007empty}. These are valleys in the pilot wave where a particle might have gone but did not, and which seemingly continue forever.

Bohmian Mechanics also does not use axiom {\bf A2}. Instead, one assumes that the only variable one can measure is the position of the Bohmian particles. Any other measurement variable that one might use in quantum mechanics must be converted into a position measurement (see e.g. \cite{durr1996bohmian}). This is an extremely important assumption of Bohmian Mechanics, because it means that all details of the measurement settings---which play a key role in Bell's notion of Local Causality---must be encoded as transformations of the state that occur before the actual measurement. The measurement settings hence become part of the Hamiltonian evolution.\footnote{One could try to argue that since the only measurement variable in Bohmian Mechanics is the position of the particles, then there is only one measurement setting. This, however, would just be a play on words, for one would still have reproduce the variables that are commonly called measurement settings in the derivation of Bell's theorem, even if one denies this is what they really are.} 

This axiomatic form of Bohmian Mechanics is not mathematically equivalent to the Copenhagen Model. One way to see this is to note that there is no equivalent to the particle positions $Q_k(t)$ in the Copenhagen Model. An easier way to see it is that if the two were equivalent, then the pilot wave should not have empty branches. However, Bohmian Mechanics in its original form is generally believed to make the same predictions as the Copenhagen Model, at least for a non-relativistic Hamiltonian (see e.g. \cite{durr1992quantum}). In the terms of \cite{taxonomy2023} then, Bohmian Mechanics is an interpretation of the Copenhagen Model, but not a representation. 

Bohmian Mechanics is also a temporally deterministic model, using the terminology of \cite{taxonomy2023}. The theory is, however, not predictive because while the initial state predicts the final state, one also cannot know the initial state exactly, hence not predict the final state. This trade-off can be done for any non-deterministic theory: By introducing hidden variables that are unmeasurable out of principle, the temporally non-deterministic time-evolution can be converted into a deterministic, yet unpredictable one. 
 
There are today many slightly different formulations of Bohmian Mechanics \cite{squires1993local,berndl1996nonlocality,durr1999hypersurface,horton2001non,dewdney2002relativistically,horton2004relativistically,durr2004bohmian,nikolic2005relativistic,nikolic2006many,colin2007dirac,struyve2010pilot,durr2014can}, and it is beyond the scope of this paper to go through all of them, but I believe they are mostly interpretations of the Copenhagen Model. An exception are those variants which do not use quantum equilibrium as one of their assumptions \cite{Valentini1991Nonequilib1,Valentini1991Nonequilib2}. Since they give under certain circumstances different predictions from the Copenhagen Model, they are modifications instead of interpretations (again using the terminology of \cite{taxonomy2023}). Those modifications will not be considered below. 

\subsection{Contextuality in Bohmian Mechanics}

 It is generally acknowledged that Bohmian Mechanics is not local because the particle velocities at one location depend on the particle positions on all other locations. This is apparent from Eq.\ (\ref{eq:Qeq}), because it implies that the dynamical evolution of the $k$-th particle depends on the position of all other particles---regardless of how far away they might be. 
 If we apply Bell's criterion of Local Causality, $P(A|S,B)=P(A|S)$ (see Figure \ref{fig:subtlety}), then Bohmian Mechanics does not fulfill it. This is because if we fully specify an initial state on $S$, then information from the measurement in $B$ will leak into the backward light cone of $A$ later via the velocity equation.

The generally accepted version of Bohmian Mechanics is non-relativistic, so one might object that we should not expect it to respect light-cones anyway. We may however note that the known (if not accepted) relativistic versions of Bohmian Mechanics still use a non-local velocity equation, see e.g. \cite{nikolic2013time}. 

Bohmian Mechanics therefore does not resolve the aspect of the measurement problem that concerns locality. While a Bohmian particle goes on one continuous path from the preparation site to the detector, the only reason this path is continuous is that it relies on a non-local guiding equation.

This explains how Bohmian Mechanics can reproduce the predictions of quantum mechanics without running afoul of Bell's theorem, but just because one assumption of Bell's theorem is violated does not mean the other ones are fulfilled. 

Notably, it has been demonstrated in \cite{hardy1996contextuality} that Bohmian Mechanics is contextual. Moreover, in \cite{Kupczynski:2022lls}, it was argued that superdeterminism---usually definied by a violation of Measurement Independence---is an example of contextuality. This might raise the impression that Bohmian Mechanics is superdeterministic, but I want to argue here that this isn't so. It rather shows that locality should be part of the definition of superdeterminism for otherwise the definition of superdeterminism is meaningless.

To illustrate this, I will use example previously proposed in \cite{bracken2021quantum}. It is a modified version of the quantum eraser experiment \cite{Scully1982Eraser,Kim2000Delayed}, in which the two different spin states of a Bell-experiment are represented as two different paths. This example has the benefit of revealing the non-locality visually because the paths are both different, whereas in the usual Bell experiments the different spins states all go on the same path.

\begin{figure}[ht]
    \centering
    \includegraphics[width=0.8\linewidth]{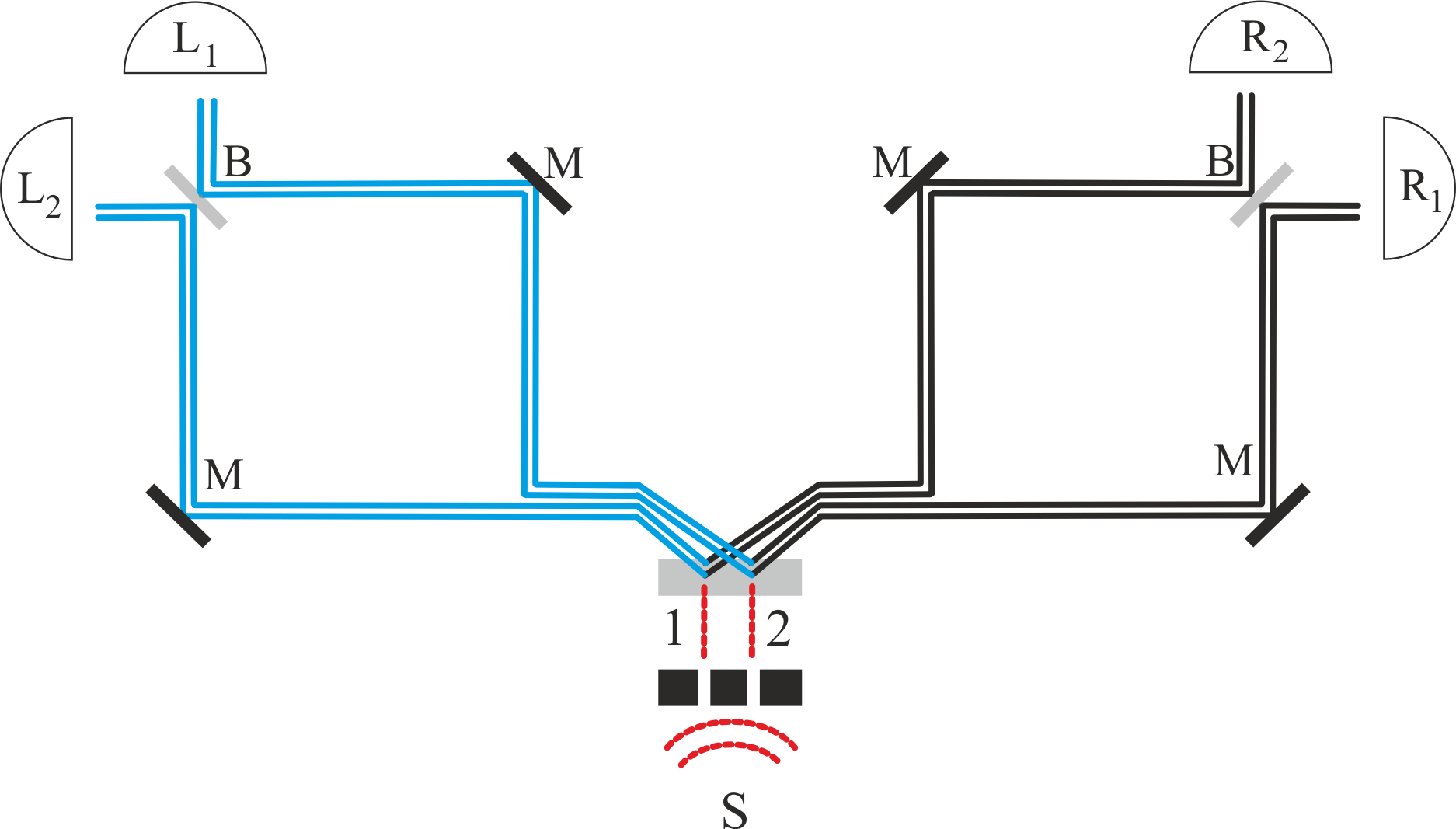}
    \caption{Sketch of the modified eraser experiment with Bohmian trajectories.}
    \label{fig:eraser01}
\end{figure}

In this experiment (Figure \ref{fig:eraser01}), photons are emitted from a single photon source (S) and sent through a double slit (black). 
After the double-slit, the photons hit a nonlinear optical crystal (light grey) which, by spontaneous parametric down conversion, creates an entangled pair of photons from each photon incident on the crystal. Note that this creates two new particles (lines in black and blue, respectively).

The photon pairs which emerge from the optical crystal are phase-matched\footnote{They are also polarisation entangled, but we do not need the polarisation in what follows.} and we denote them with $|{\rm 1}\rangle = (1,0)^{\rm T}$ and $|{\rm 2}\rangle = (0,1)^{\rm T}$, depending on which slit they came from. The photons on each side then each enter a Mach-Zehnder interferometer lacking the first beam splitter, with paths adjusted so that photon counts are the same in L$_1$ and L$_2$, or R$_1$ and R$_2$, respectively. 

In this setup, the location at which the entangled particles are created (1 or 2) imprints the information that is later erased when the states are mixed at the beam splitters. The erasure then comes down to the somewhat unremarkable statement that if we measure one of the entangled particles in L$_1$ (L$_2$), then its partner goes to R$_1$ (R$_2$). 

But on both paths of the entangled photons we can alternatively chose to measure the which path information with detectors we will call R$_{3/4}$ and L$_{3/4}$ (not shown in Figure). If we measure a combination of L$_{1/2}$ and R$_{3/4}$ (or the other way round), they should not be correlated at all.

We adopt the convention that the first entry in the product state denotes the photon going left, the second the photon going right. Then the state after the crystal can be written as
\begin{eqnarray}
|\Psi \rangle &=& \frac{1}{\sqrt{2}} \Big( |{\rm 1}\rangle_{\rm L} |{\rm 1}\rangle_{\rm R} + | {\rm 2}\rangle_{\rm L} | {\rm 2}\rangle_{\rm R} \Big) \\
&=& \frac{1}{\sqrt{2}} \Big( |-\rangle_{\rm L} |-\rangle_{\rm R} + |+\rangle_{\rm L} |+\rangle_{\rm R} \Big) ~. \label{eq:psi}
\end{eqnarray} 
where 
\begin{eqnarray}
 |\pm \rangle_{\rm L/R}  := \frac{1}{\sqrt{2}} \left(
 |1 \rangle_{\rm L/R} 
\pm |2 \rangle_{\rm L/R} 
\right)~.
\end{eqnarray}

The actions of the which-path detectors are
\begin{eqnarray}
 {\rm L}_3:  |1 \rangle_{\rm L}  \langle  {\rm 1}|_{\rm L}  \otimes {\rm \mathds{1}}_{\rm R}  ~,~
 {\rm L}_4:  |{\rm 2} \rangle_{\rm L}   \langle {\rm 2} |_{\rm L}  \otimes {\rm \mathds{1}}_{\rm R}   ~, \nonumber \\ 
 {\rm R}_3: {\rm \mathds{1}}_{\rm L}  \otimes |1\rangle_{\rm R} \langle  {1}|_{\rm R}  ~,~
 {\rm R}_4:  {\rm \mathds{1}}_{\rm L}  \otimes |2 \rangle_{\rm R}   \langle {\rm 2} |_{\rm R}   ~. \end{eqnarray}
The detectors L$_{1/2}$ and R$_{1/2}$ instead project on the bases rotated by the action of the beam splitter $B$
so that
\begin{eqnarray}
{\rm L}_1:    | - \rangle_{\rm L}  \langle -|_{\rm L}  \otimes {\rm \mathds{1}}_{\rm R} ~,~
 {\rm L}_2:   |{\rm +} \rangle_{\rm L}   \langle {\rm +} |_{\rm L} \otimes {\rm \mathds{1}}_{\rm R}    ~, \nonumber \\
{\rm R}_1: {\rm \mathds{1}}_{\rm L}  \otimes |{\rm -} \rangle_{\rm R}  \langle  {\rm -}|_{\rm R}  ~,~
 {\rm R}_2:  {\rm \mathds{1}}_{\rm L}  \otimes |{\rm +} \rangle_{\rm R}   \langle {\rm +} |_{\rm R}    ~. 
\end{eqnarray}

We now take into account, as pointed out in \cite{hardy1996contextuality}, that the evolution law for the Bohmian particles Eq.\ (\ref{eq:Qeq}) is a first order differential equation in time. For this reason, the possible trajectories of one Bohmian particle cannot cross. This is similar to how trajectories in phase-space cannot cross. Trajectories of Bohmian particles belonging to different photons can cross because they obey different differential equations. 

Since the trajectories of Bohmian particles cannot cross, the particles that arrive at a beam splitter first are the only ones that can go through. If one is reflected, all later ones also have to be reflected. When recombining the beams in a non-interference situation, the beam splitter cannot mix paths---again this is because the trajectories cannot cross. 

If one uses these requirements, one finds that if we recombine the paths on each side and measure interference, then the only possible way the Bohmian particles can go is as illustrated in Fig. \ref{fig:eraser01}. However, if we make a which path measurement on the right side rather than recombining the beams and post-select, then one of the trajectories must go through the beam splitter on the left side, as shown in Fig. \ref{fig:eraser01}. That is, the path of one of the Bohmian particle trajectories that we have drawn depends on the measurement setting. Note that the measurement on the right side could be done after the measurement on the left side (time-like separated), so this cannot be explained by entanglement with the detector.  

\begin{figure}[ht]
    \centering
    \includegraphics[width=0.8\linewidth]{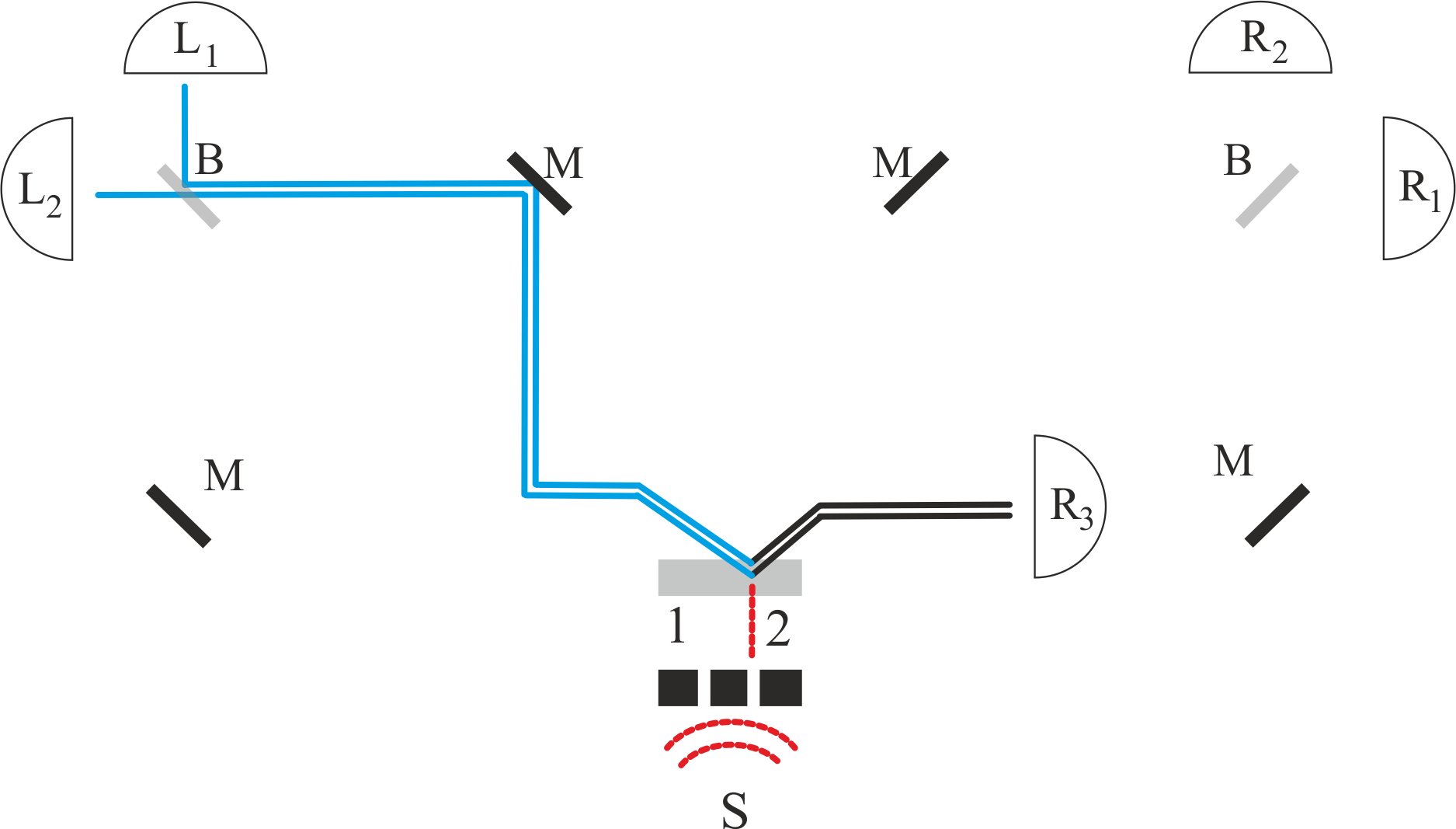}
    \caption{Sketch to illustrate how a which-way measurement on the right side must change the Bohmian trajectories on the left side, which could be causally disconnected.}
    \label{fig:eraser02}
\end{figure}

If we generally denote hidden variables with $\lambda$ and the detector settings (of possibly several detectors) with $X$, then a violation of Measurement Independence means $\rho(\lambda|X) \neq \rho(\lambda)$, where $\rho$ is the probability density of the hidden variables. This correlation should not be evaluated after or during the measurement, because once the state that one measures is in contact with the measurement apparatus, the hidden variables will of course become correlated with the measurement settings. A violation of Measurement Independence means that the hidden variables and the measurement settings were correlated already when the measurement begins. In the above example, Measurement Independence is violated because the hidden variable labeling the path going to L$_2$ in Fig \ref{fig:eraser02} must change depending on the measurement variable. 

However, the violation of Measurement Independence comes about because Bohmian Mechanics is non-local. The paths of particles on the right side can get information about what their entangled partners on the left side did, immediately.  Though it is eventually a matter of nomenclature, for me this example illustrates that it makes no sense to define superdeterminism by merely a lack of Measurement Independence. One also needs to require locality, for non-locality can induce violations of Measurement Independence. It would be confusing to lump Bohmian Mechanics together with models that employ violations of Measurement Independence to restore Local Causality.

This example is also useful to address another common misunderstanding, which is that a violation of Measurement Independence in a hidden variables theory is a statement about the initial conditions for the hidden variables. 

This probability density is usually interpreted as one at the time of preparation of the system (see e.g. \cite{Pusey2012Reality}). However, 
as pointed out in \cite{hance2022ensemble}, the quantity $\rho(\lambda)$ appears in Bell's theorem to calculate the measurement result. Initial values for the hidden variables without an evolution law are not sufficient for that. Hence, the $\lambda$s in Bell's theorem must necessarily include information about the evolution law and Measurement Independence can be violated by the evolution law, rather than by the initial condition. This is exactly what happens in Bohmian Mechanics.

The initial condition of the particles in Bohmian mechanics does not depend on the measurement settings as it is just a way of re-expressing the initial state of the wave-function in the Copenhagen Model. But since in Bohmian Mechanics the evolution law mediates non-local influence, the correlation can be created during the evolution of the state. 

\subsection{Measurements in Bohmian Mechanics}

Let us then look at what Bohmian Mechanics has to say about the measurement problem.

We already saw that Bohmian Mechanics is non-local, so it does not solve that aspect of the measurement problem. But what about the other aspect of the measurement problem, that which explains what makes a collection of particles act as a measurement device? Well, in Bohmian Mechanics there is no need for an apparatus to bring a state into an eigenstate of the operator that describes the measurement variable because the only thing that can be measured in in this model is the position of the Bohmian particles, and this position always has a definite value anyway. That is, this part of the measurement problem doesn't exist in Bohmian Mechanics for the same reason it doesn't exist in classical mechanics. 

In a classical theory we have no problem with explaining why, say, a thermometer tells us the temperature of a system, because that's just a consequence of the interaction between the thermometer and the system we are interested in. This works because the system has a temperature whether or not we measure it and we just need to read it out. True, we often don't write down this interaction between the measurement device and the system in detail, but we know how it works in principle.\footnote{We may note in the passing that even in classical mechanics, the measurement device necessarily affects the system that it is measuring because the two need to interact. This is not an exclusively quantum phenomenon.}

This is not the case in the Copenhagen Model, where the prepared state a priori simply does not have a definite value for the variable we want to measure (because it's in a superposition), and the interaction between the system and the measurement device is a unitary transformation that just maintains those superpositions. It is for this reason that, in the Copenhagen Model, the measurement device needs to somehow make the measurement variable definite by ``collapsing'' superpositions. But in Bohmian Mechanics this is not necessary. It is therefore not so much that Bohmian Mechanics solves the problem as that it doesn't have it to begin with.

The measurement process in Bohmian mechanics itself works similar to decoherence, in that one regards the system of prepared state plus the detector and the environment.\footnote{It is worth mentioning that Bohm's elaborations on the measurement process in Bohmian Mechanics predates much of the later work on decoherence.} Then one effectively removes the environment and detector by creating what is called the ``conditional wave-function'', by inserting the positions of the Bohmian particles that one is not interested in. The remaining wave-function then effectively contracts the probability distribution which can be interpreted as a sort of collapse. 
However, much like in decoherence approaches, this idea requires one to know already that a collection of particles will act as a measuring device. 

So does Bohmian Mechanics solve the part of the measurement problem that pertains the identification of what constitutes a detector? It does so as much or as little as decoherence and the Many Worlds Interpretation do. That is, if one designates part of the state as detector and environment, then one can create a conditional wave-function for the prepared state that evolves into a position eigenstate, but one does not know from the initial state alone what constitutes the detector and its environment. 

\section{Discussion}
\label{sec:Disc}
 
As always there is more that could be said. For example, to me the above discussion raises the question of whether temporal determinism is even a meaningful criterion, given how easily it can be re-established by positing mathematical entities (be that disconnected parallel worlds or particles with unknowable positions) that are unmeasurable, even in principle. Another interesting question is why Bohmian Mechanics needs to violate Local Causality when it also violates Measurement Independence. It seems to me that either it doesn't actually violate Local Causality, or it must be possible to find a modified, yet empirically equivalent, version that does respect Local Causality.

Finally, I would like to submit that I have no personal preference for either of these interpretations of quantum mechanics. To me they are all equally useful and equally unsatisfactory. I merely hope that this modest contribution to the literature will help identify just what is unsatisfactory about them.

\section{Summary}
\label{sec:Sum}
 
In summary, I have applied an instrumentalist approach to explain why both the Many Worlds Interpretation and Bohmian Mechanics are non-local and solve the measurement problem only partially, just like the Copenhagen Interpretation.

\section*{Acknowledgements}

I thank Emily Adlam, Jonte Hance, and Hrvoje Nikolic for helpful correspondence.

\bibliographystyle{unsrturl}
\bibliography{ref.bib}

\end{document}